\documentclass{aa} 
\usepackage[varg]{txfonts}
\usepackage{graphicx}
\usepackage{txfonts}
\usepackage{natbib}
\usepackage{epsfig}
\usepackage{deluxetable}
\usepackage{longtable}

\def\simgt{\lower.5ex\hbox{\gtsima}}

\newcommand{\fgas}{$f_{\rm gas}$}

\newcommand{\s}{{\it Spitzer}}
\newcommand{\h}{{\it Herschel}}
\newcommand{\msol}{$\rm M_{\odot}$}
\newcommand{\md}{$M_{\rm dust}$}

\newcommand{\lsol}{L$_{\odot}$}
\newcommand{\lir}{$L_{\rm IR}$}
\newcommand{\ms}{$M_{\ast}$}
\newcommand{\lco}{$L^{\prime}_{\rm CO}$}

\newcommand{\mol}{$M_{\rm H_{\rm 2}}$}
\newcommand{\td}{$T_{\rm d}$}
\newcommand{\aco}{$\alpha_{\rm CO}$}

\newcommand{\mgas}{$M_{\rm gas}$}

\newcommand{\cooz}{CO[J=1$\rightarrow$0]}
\newcommand{\coto}{CO[J=2$\rightarrow$1]}
\newcommand{\cott}{CO[J=3$\rightarrow$2]}
\newcommand{\coyy}{CO[J=4$\rightarrow$5]}

\newcommand{\Mstar}{$M_{\ast}$}
\newcommand{\um}{$\langle U \rangle$}
\title{Dust and Gas in Star Forming  Galaxies at $z\sim3$ - \\Extending Galaxy Uniformity to 11.5 Billion Years.
\thanks{Herschel is an ESA space observatory with science instruments provided
  by European-led Principal Investigator consortia and with important participation from NASA.}}

\author{Georgios E. Magdis\inst{1,2}
\and D.~Rigopoulou \inst{3,4}
\and E. Daddi\inst{5}
\and M. Bethermin\inst{6}
\and C. Feruglio\inst{7}
\and M. Sargent\inst{8}
\and H. Dannerbauer\inst{9,10}
\and M. Dickinson\inst{11}
\and D. Elbaz\inst{5}
\and C. Gomez Guijarro\inst{1}
\and J-S Huang\inst{12,13,14}
\and S. Toft\inst{1}
\and F. Valentino\inst{1}
}
\institute{Dark Cosmology Centre, Niels Bohr Institute, University of Copenhagen, Juliane Mariesvej 30, DK-2100 Copenhagen, Denmark \\
\email:{magdis@dark-cosmology.dk}
\and Institute for Astronomy, Astrophysics, Space Applications and Remote Sensing, National Observatory of Athens, GR-15236 Athens, Greece
\and Department of Physics, University of Oxford, Keble Road, Oxford OX1 3RH, UK
\and Space Science \& Technology Department, Rutherford Appleton Laboratory, Chilton, Didcot, Oxfordshire OX11 0QX, UK
\and CEA, Laboratoire AIM, Irfu/SAp, F-91191 Gif-sur-Yvette, France
\and Aix.  Marseille  Univ,   CNRS,  LAM,  Laboratoire  d'Astrophysique  de  Marseille,   Marseille
\and INAF Observatory of Trieste - Via G.B. Tiepolo 11, Trieste - Italy
\and Astronomy Centre, Department of Physics and Astronomy, University of Sussex, Falmer, Brighton BN1 9QH
\and Instituto de Astrofisica de Canarias (IAC), E-38205 La Laguna, Tenerife, Spain  
\and Universidad de La Laguna, Dpto. Astrofisica, E-38206 La Laguna, Tenerife, Spain
\and NOAO, 950 N. Cherry Avenue, Tucson, AZ 85719, USA
\and National Astronomical Observatories of China, Chinese Academy of Sciences, Beijing 100012, China
\and China-Chile Joint Center for Astronomy, Chinese Academy of Sciences, Camino El Observatorio, 1515, Las Condes, Santiago, Chile
\and Harvard-Smithsonian Center for Astrophysics, 60 Garden Street, Cambridge, MA 02138, USA)
}
\begin{document}
\abstract{We present millimetre dust emission measurements of two Lyman Break Galaxies at $z\sim3$ and construct for the first time fully sampled infrared spectral energy distributions (SEDs), from mid-IR to the Rayleigh-Jeans tail, of individually detected, unlensed, UV-selected, main sequence (MS) galaxies at $z=3$. The SED modelling of the two sources confirms previous findings, based on stacked ensembles, of an increasing mean radiation field \um\ with redshift, consistent with a rapidly decreasing gas metallicity in $z > 2$ galaxies. Complementing our study with \cott\ emission line observations, we measure the molecular gas mass reservoir (\mol) of the systems using three independent approaches: 1) CO line observations, 2) the dust to gas mass ratio vs metallicity relation and 3) a single band, dust emission flux on the Rayleigh-Jeans side of the SED. All techniques return consistent \mol\ estimates within a factor of $\sim$2 or less, yielding gas depletion time-scales ($\tau_{\rm dep} \approx 0.35$\,Gyrs) and gas-to-stellar mass ratios (\mol/$M_{\ast} \approx 0.5-1$) for our $z \sim3 $ massive MS galaxies. The overall properties of our galaxies are consistent with trends and relations established at lower redshifts, extending the apparent uniformity of star forming-galaxies over the last 11.5 billion years.}

\titlerunning{Dust and Gas in LBGs at $z\sim3$}
\authorrunning{G. E. Magdis et al.}

\maketitle
\section{Introduction}
Star formation in galaxies proceeds through the conversion of molecular hydrogen into stars, in dense molecular clouds (e.g.,  Fukui  \&  Kawamura 2010). It naturally 
follows that any attempt to understand galaxy evolution and characterise the star formation history of the Universe requires the measurement of the molecular hydrogen mass reservoir (\mol) of galaxies across cosmic time. To this end, several techniques that convert direct observables, or well calibrated derived quantities, to \mol, have been developed and successfully applied. Applying these techniques to galaxy populations that meet various selection criteria and lie at different cosmic epochs, has revealed a consistent picture where the star formation rate and the molecular gas mass of the majority of star forming galaxies at any redshift, are tightly correlated, following the so-called Schmidt-Kennicutt relation. This along with the discovery of the main sequence of star formation (MS) approximately a decade ago (e.g., Noeske et al.\,2007, Elbaz et al.\,2007), provide evidence of a uniformity in the star formation histories and the star formation activity of the galaxies at least up to $z \approx 2$ (e.g. Daddi et al.\,2010, Magdis et al.\,2012b, Sargent et al.\,2014, Genzel et al.\,2015, Tacconi et al.\,2017). 

Measuring  \mol\  in galaxies though,  becomes progressively more difficult and more demanding, in terms of required observational time per object, with look-back time. For example, CO lines,  that are the most traditional gas mass tracer, have only been measured in a very small fraction of spectroscopically confirmed $z>1$ galaxies (e.g. Solomon \& Vanden Bout\,2005; Carilli \& Walter\,2013, Bothwell et al.\,2013, Aravena et al.\,2016, Silverman et al.\,2015 Tacconi et al.\,2017), while at $z > 2.5$ are primarily restricted to lensed objects or strong starbursts (e.g. Saintonge et al.\,2013, Dessauges-Zavadsky et al.\,2015). Similarly, the dust to gas mass technique ($\delta_{\rm GD}$), that has become increasingly popular at high$-z$ thanks to the advent of the Herschel Space Observatory (\h), requires detailed sampling of the far-IR SED and of the Rayleigh-Jeans (R-J) tail of the SED, and therefore is primarily restricted to $z<2$ or extremely bright and rare sources at higher redshifts, due to the limited sensitivity of \h. Thus, the gas fraction and star formation efficiency of ``normal" (MS) galaxies remain poorly constrained at $z>2$. 

In an attempt to extend \mol\ studies of normal galaxies to higher redshifts,  Bethermin et al.\,(2015), applied the $\delta_{\rm GD}$ method in stacked ensembles of galaxies up to $z=4$ (see also Santini et al.\,2014, Genzel et al.\,2015 and Tacconi et al.\,2017). Also, Scoville et al.\,(2017) and Schinnerer et al.\,(2016), were able to place constraints on the  gas masses of individual galaxies up to $z=4$, using single band measurements of the dust emission flux on the Rayleigh-Jeans side of their SED. Finally, Magdis et al.\,(2012a), and Tan et al.\,(2014), presented CO observations, and therefore \mol\ estimates for a couple of $z=3-4$  Lyman Break Galaxies (LBGs). Clearly, we need a larger  sample of high$-z$ ``normal" galaxies for which more than one method to derive \mol\ estimates can be applied, in order to increase the currently limited number of $z>2.5$ galaxies with measured \mol\ but also check against systematics in \mol\ estimates  among the various methods. 

In this paper we combine mid to far-IR data with millimeter observations of the R-J tail and \cott\ emission line detections of two $z\sim3$, UV selected, Lyman Break galaxies that lie on the MS of star formation, in order to 1) study their far-IR properties, 2) derive gas mass estimates using three independent techniques and 3) investigate the gas depletion time scales and gas fraction of individual MS galaxies at $z=3$. Throughout the paper we refer to the molecular hydrogen gas mass as \mol\ and to the total gas mass, which is the sum of \mol\ and of the atomic gas mass ($M_{\rm HI}$), as \mgas. We adopt $\Omega_{\rm m}$ = 0.3, $H_{\rm 0}$ = 71 km s$^{-1}$ Mpc$^{-1}$, $\Omega_{\rm \Lambda}$ = 0.7 and a Chabrier IMF (Chabrier et al.\,2007).
 
\section{Sample and observations}
 \begin{figure}
\centering
\includegraphics[scale=0.16]{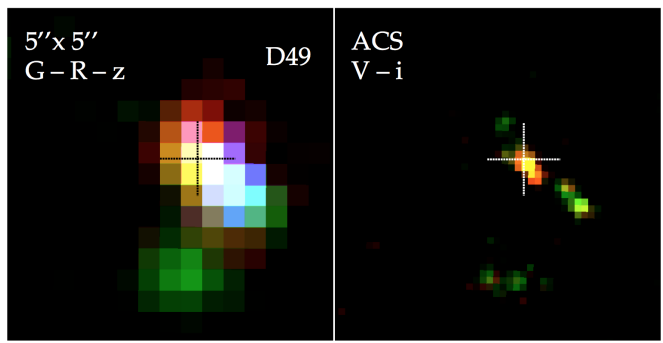}
\caption{\textbf{Left:} G-R-z, three colour image of D49, as obtained by ground based imaging. There is a clear colour gradient with CO emission centered at the position of the red component. \textbf{Right:} V-i, high-resolution two colour image obtained by ACS/HST. The system is resolved into two components ( a red and a blue), separated by $\sim$1''.  The cross (and its size) in each panel depicts the centroid (and the position uncertainty) of the line emission detected by PdBI.} 
\label{fig:image} %
\end{figure}

\subsection{Optical to mid-IR data}
The two galaxies of this study, D49 (R.A.: 214.37169, decl.: +52.576303) and M28 (R.A.: 214.44282, decl.: $+52.45561$), were drawn from the optically selected ($U,G,R$) sample of  $z \sim 3$ Lyman Break Galaxies by Steidel et al.\,(2003) in  the Extended Groth Strip field (EGS). Their selection was based on the available spectroscopic redshift ($z = 2.808$ and $z = 2.903$)   determined by Ly-$\alpha$ emission through ground based rest-frame UV spectroscopy (Steidel et al.\,2003, Shapley et al.\,2003). Since the galaxies were selected for CO and dust continuum follow-up observations, a second criterion was a detection at 24\,$\mu$m to ensure sufficiently high total infrared luminosities (\lir\ $>$ 10$^{12}$\,\lsol) that would make the detection experiment possible (Rigopoulou et al. 2010, Magdis et al. 2010b). This criterion biases our sample to the most massive and infrared luminous LBGs at $z \sim 3$ (e.g Rigopoulou et al. 2006, Magdis et al. 2010b), which however have overall properties (in terms of specific star formation rate and star formation efficiency) similar to that of MS galaxies (see sections 3 and 4). Our sources benefit from extensive multi-wavelength coverage  (Rigopoulou et al.\,2006)  including ground based  (U,G,R,J,K,z) observations and photometry from the Advanced Camera for Surveys (ACS, F606W (V), F814W (i)), Infrared Array Camera (IRAC), and the Multi-band Imaging Photometer (MIPS) on board the Spitzer Space Telescope (\s). 

\subsection{\h\ data}
The EGS field has been observed by the Herschel Space Observatory (\h) as part of the PACS Evolutionary Probe (PEP; Lutz et al.\,2011) and the Herschel Multi-tiered Extragalactic Survey (HerMES, Oliver et al.\,2010, 2012), providing  100 and 160\,$\mu$m imaging with the Photodetector Array Camera and Spectrometer (PACS; Poglitsch et al.\,2010) and 250, 350, and 500\,$\mu$m with the Spectral and Photometric Imaging Receiver (SPIRE; Griffin et al.\,2010), respectively. Since both our sources are detected at 24\,$\mu$m, we performed \h\ photometry by using the source extraction point-spread function fitting code {\textit{Galfit}} (Peng et al. 2002), guided by 24\,$\mu$m priors. As a sanity check, we also compared the derived fluxes of our sources with those reported in the photometric catalogues provided by the PEP and HerMES teams. The catalogues were also produced based on  a 24\,$\mu$m prior source extraction technique. The derived fluxes between the various catalogues  are in excellent agreement yielding a S/N $>$ 3 detection of D49 in all 5 \h\ bands and of  M28  at 250- and 350\,$\mu$m. The derived \h\ fluxes for the two sources are summarised in Table 1.

\subsection{Millimetre observations}
The 1.2\,mm continuum observations presented here were carried out during 
December 2005 and February 2006 using the
117-channel Max-Planck Millimetre Bolometer (MAMBO-2, Kreysa et al.\,1999)
array at the Institut de Radioastronomie Millimetrique (IRAM) 30m 
telescope on Pico Veleta (Spain).
MAMBO-2 operated at an effective wavelength of 1.2\,mm which corresponds to 250\,GHz (FWHM $\sim$10''.7). The
sources were observed with the array's central channel, using the
standard on-off mode with the telescope secondary chopping in azimuth
by 50'' at a rate of 2\,Hz. The target was always positioned on the
central bolometer of the array and after 10\,s of integration the
telescope was nodded so that the off-beam became the on-beam. 
On-off observations were obtained in 20\,mins scans with a wobbler throw 
of 35'' and were repeated until either a detection (S$/$N $\geq$ 3) or a
noise level of $\sim$0.4\,mJy was reached.
Pointing was checked frequently
on nearby continuum sources, and was found on average stable to
2''. Sky opacity was monitored regularly with zenith opacities
at 1.2\,mm and was found to vary between 0.06 and 0.4. Primary pointing, focus
and flux calibrations were performed using observations of Mars ($\sim$130\,mJy) or CW-LEO ($\sim$1.3\,mJy). 
 
The data were analysed using MOPSIC, an updated version of the MOPSI software developed by R. Zylka. 
Every scan and its associated sub-scans were carefully
inspected for outliers or influence of high opacity -- affected sub-scans were
subsequently excluded from further reduction.  Correlated sky-noise was computed for each channel separately
as a weighted mean of the signals from the surrounding channels. Then the mean was subtracted from each channel. Both sources are detected at S/N $>$ 3, with $S_{\rm 1.2mm} = 1.76 \pm 0.37$\,mJy for D49 and $S_{\rm 1.2mm} = 1.10 \pm 0.36$\,mJy (Table 1).
 \begin{figure}
\centering
\includegraphics[scale=0.29]{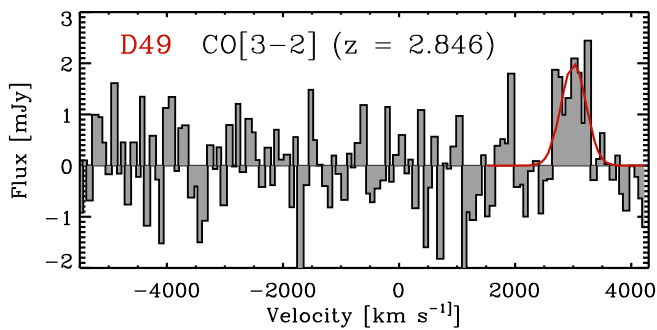}
\caption{Spectrum of \cott\ emission of D49, binned in steps of 65\,km\,s$^{-1}$. The detected line is offset by $\sim$\,3000\,km\,s$^{-1}$, with respect to the central frequency expected based on the redshift of the source obtained by Steidel et al.\,(2003). The red line corresponds to the gaussian fit to the detected line profile.}
\label{fig:spec} %
\end{figure}
 \begin{figure*}[!h]
\centering
\includegraphics[scale=0.26]{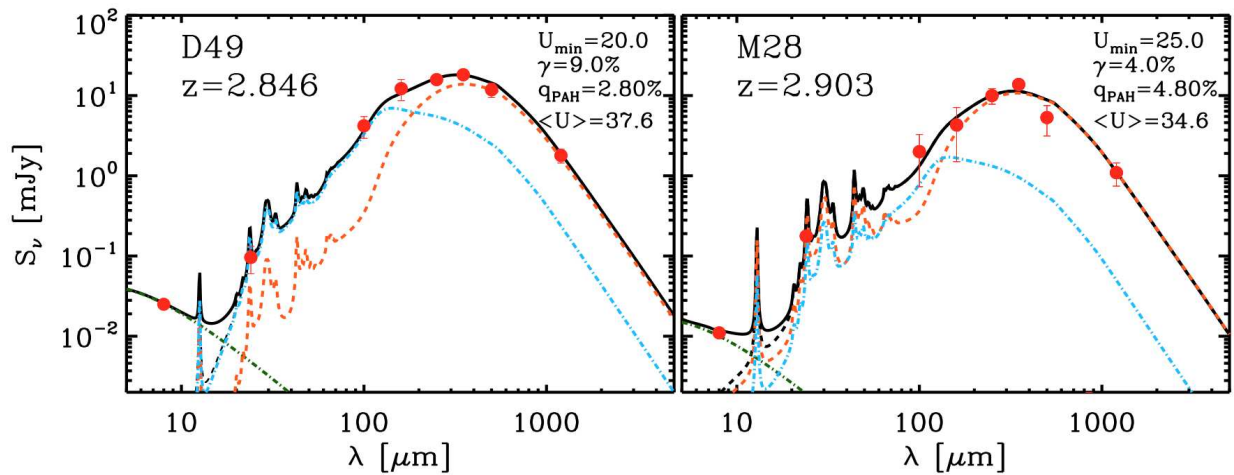}
\caption{Spectral energy distributions (observed frame) of D49 and M28, including IRAC 8\,$\mu$m, MIPS 24\,$\mu$m PACS 100, 160\,$\mu$m and SPIRE 250, 350, and 500\,$\mu$m and MAMBO 1.2\,mm data points (red dots). The observed data are overlaid with the best-fit DL07 model (black line). The ``PDR" and diffuse ISM components are shown in cyan and orange, respectively, while the stellar component is depicted with a green dashed-dotted line. On the top right of each panel we list the derived parameters from the SED modeling, including $U_{\rm min}$ (radiation field of the diffuse ISM), $\gamma$ (the contribution of the PDR component in the total SED), $q_{\rm PAH}$ (the fraction of dust in PAHs) and $\langle$U$\rangle$ (mean radiation field).}
\label{fig:seds} %
\end{figure*}
\subsection{PdBI observations of D49}
We targeted the \cott\ ($\nu_{\rm rest}$ $=$ 345.796\,GHz) transition towards  D49, using the Plateau de Bure Interferometer. At $z = 2.808$ this line is redshifted to 90.8078\,GHz. Observations were carried out under excellent 3\,mm weather conditions in D array 
configuration with five and six antennas for 2.77 and 1.85\,hrs respectively, in June 2014. Data reduction was performed using CLIC in GILDAS. The noise level reach by the combined observations is 0.29\,mJy/beam in 128\,MHz channels ($\sim$\,400\,km\,s$^{-1}$) and the FWHM of the circularized synthesized beam is about 6''.  We note that CO observations of M28 were not carried out. 

We searched for CO emission at the optical position of the LBG (R.A.: 214.37169, decl.: +52.576303) and detected an emission line at S/N $=$ 5.8 (Fig. \ref{fig:spec}). The spatial position of the peak line emission (R.A.: 214.37181, decl.: +52.576236 $\pm 0.6''$) is consistent with the optical counterpart within the uncertainties  but the detected line is redshifted $\sim$\,3000\,km\,s$^{-1}$ with respect to the expected frequency based on the optical redshift of the source. The fact that Ly-$\alpha$ is blueshifted rather than redshifted with respect to CO,  does not support the scenario where the discrepancy originates from an extinction truncated Ly-$\alpha$ profile. Also, the scenario of outflows due to the presence of a strong AGN is disfavoured by the absence of strong AGN activity, advocated by the rest-frame UV spectrum, the lack of detection in the X-rays and the IRAC colours (e.g. Donley et al.\,2012). On the other hand, a possible solution is that that the Ly-$\alpha$ and the 3\,mm emission line are arising from two different but nearby systems; a dust obscured one and relatively dust free Ly-$\alpha$ emitting component. 

This is supported from the 3 colour image (GRz) shown in (Fig. \ref{fig:image} left), that reveals a colour gradient within the single source detected by ground based imaging. Upon inspection of the ACS images, that were not available in the original study of Steidel et al.\,(2003), it becomes evident that the single source seen in the ground based imaging is deblended into two components.  Again, from the two colour ACS image (V-i, Fig. \ref{fig:image} right) it is clear that the system originally selected as a single LBG, consists of two components, a blue and a red that are $\sim$1'' apart. We note that the IRAC, MIPS 24\,$\mu$m as well as the line emission are centered at the position of the red component. Assuming that the detected line is \cott, the observed central frequency yields are redshift of $z=2.846$ for the red component. Using the ACS photometry, and $\lambda$ > 6000\,$\AA$ ground based photometry  (where the bulk of the emission  is originating from the red component), we run the Le-Phare and Hyper-z phot$-z$ codes to infer a photometric redshift for the red component. Both codes return a photo$-z$ of 2.95 $\pm$ 0.22. Similarly the mid-IR to mm SED is consistent with a galaxy at $z \sim3 $ (see section 4). An alternative scenario where the detected line is \coyy\  of a $z = 4.11$ source, can be ruled out from the photo$-z$ analysis as well as from the shape of the infrared SED.  We conclude that detected line is \cott, originating from the red component that lies at $z_{\rm CO} = 2.846 \pm 0.011$, and which is responsible for the bulk  of the near-IR to mm emission. We note that the CO redshift is inconsistent by more than 3$\sigma$ with respect to the optical redshift derived by fitting the Ly-$\alpha$ emission line from the UV spectrum ($z_{\rm opt} = 2.8079 \pm 0.0003$). Finally, we stress that  whether this a single system consisting of two nearby galaxies lying 8\,kpc apart (as indicated by their angular distance) and moving with 3000\,km\,s$^{-1}$ with respect to each other, or whether the two sources lie $\sim 40$\,Mpc apart (as indicated by their redshifts described above), does not change the results presented in this paper, as the bulk of CO, dust and stellar ($>$90\%) emission originates from the red component.  
\section{Results}
\subsection{Far-IR properties}
The far-IR properties of $z\sim3$ LBGs have been investigated in previous studies (e.g. Magdis et al.\,2010b, Rigopoulou et. al.\,2010, Magdis et al.\,2012a, Coppin et al.\,2015). However, the majority of these studies are based on stacking results or on sparse sampling of the full IR SED.  Here, we present and model mid-IR to mm SEDs of individually detected LBGs, with detailed coverage of the peak and the R-J part of the spectrum (Fig. \ref{fig:seds}).  

We combine \s\ IRAC (8.0\,$\mu$m), MIPS (24\,$\mu$m), \h\ PACS/SPIRE  (100, 160 250, 350, and 500\,$\mu$m) with the 1.2mm IRAM continuum measurements and use the Draine \& Li (2007, hereafter DL07) model to derive dust masses and infrared luminosities by fitting the mid-IR to sub-millimeter photometry. The analysis is also supplemented by a more simplistic but widely used single temperature modified blackbody (MBB) fit in order to derive a representative single dust temperature \td\ of the ISM. In particular, we adopt a fixed effective dust emissivity index of $\beta = 1.8$ and fit observed data points with $\lambda_{\rm rest} > $ 50\,$\mu$m to avoid emission from very small grains that dominate at shorter wavelengths. A similar technique for the derivation of \md, \td , and \lir\ using the DL07 and MBB models is presented in detail in Magdis et al. (2012b, 2013). The best-fit model SEDs are presented in Fig. \ref{fig:seds}.  The derived luminosities and dust masses are, \lir\ $= (6.02 \pm 0.41) \times 10^{12}$\,\lsol\ and \md\  $=  (1.34 \pm 0.36) \times 10^{9}$\,\msol\  for D49 and \lir\ $= (3.16 \pm 0.29) \times 10^{12}$\,\lsol\ and \md\ $=  (7.95 \pm 2.74) 10^{8}$\,\msol\  for M28. Similarly, the corresponding dust temperatures obtained from the MBB model  are 41 $\pm$ 2\,K and  41.8 $\pm$ 3\,K for D49 and M28 respectively. The derived far-IR properties of the galaxies  are summarised in Table 2.

\subsection{CO emission properties}
To measure the velocity centroid and line width of the CO emission from D49, we fit a single Gaussian to the observed spectrum (Fig. \ref{fig:spec}), extracted at the  phase center, that coincides with the position of the red component seen in the optical band. The fit yields a peak flux density of $S_{\rm CO} =$1.99 $\pm$0.12\,mJy, a FWHM of 501 $\pm$ 35\,km\,s$^{-1}$ and the  velocity integrated flux is $I_{\rm CO}$ = 1.10 $\pm$ 0.18\,Jy\,km\,s$^{-1}$. We derive the  \cott line luminosity of D49 ($L_{\rm CO[3-2]}$, in K\,km\,s$^{-1}$\,pc$^{2}$),  from the detected velocity integrated flux ($S_{\rm CO} \Delta u$) using  the eq. (3) in Solomon \& Vanden Bout\,(2005) :

\begin{equation}
\begin{centering}
L_{\rm CO[3-2]} =  3.25 \times 10^{7}\,S_{\rm CO} \Delta u\, \nu^{-2}_{\rm obs}\,D^{2}_{\rm L}\,(1+z)^{-3}
\end{centering}
\end{equation}

\noindent where $S_{\rm CO} \Delta u$ is in Jy\,km\,$\rm s^{-1}$, $\nu_{\rm obs}$ is the observed frequency of the detected  line (\cott) in GHz, and $D_{\rm L}$ is the luminosity distance in Mpc. The inferred \cott\, luminosity is $L_{\rm CO[3-2]} = (4.16 \pm 0.68) \times 10^{10}$\,K\,km\,s$^{-1}$\,pc$^{2}$.

\subsection{Position relative to Main Sequence}
\begin{figure}[!h]
\centering
\includegraphics[scale=0.4]{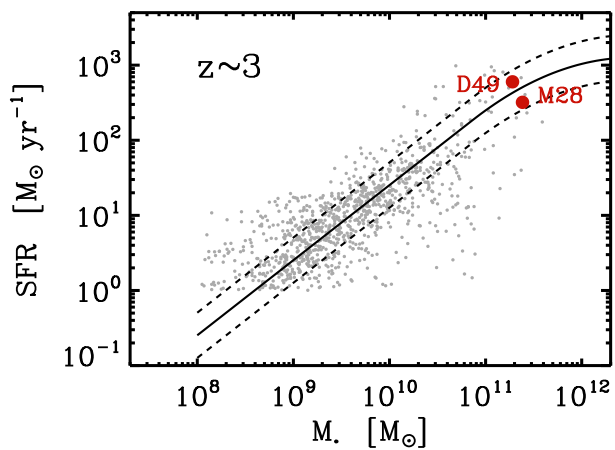}
\caption{Location of D49 and M28 (red circles) with respect to the main sequence at $z=3$ (solid line) and its scatter (dashed lines), as measured by Schreiber et al. (2015). Grey circles correspond to $2.7<z<3.2$ galaxies in GOODS-S with measured SFR and \Mstar.}
\label{fig:ms} %
\end{figure}

To determine the position of the two galaxies with respect to the MS, we need to determine the SFRs and stellar masses. For the former, we convert the inferred IR luminosities 
to SFRs adopting the Kennicutt\,1998 SFR to \lir\ conversion for a Chabrier IMF. For the latter, we consider the available UV to near-IR photometry and use the BC03 models 
assuming a constant star formation history. Our analysis yields a star formation rate of 600\,\msol\,yr$^{-1}$ and a stellar mass of 1.9 $\times 10^{11}$\,\msol\ for D49 while the 
corresponding values for M28 are SFR $=$ 320 \msol\,yr$^{-1}$ and $M_{\ast} = 2.4 \times 10^{11}$\,\msol. We note that the inferred stellar masses are in agreement with the 
values reported in Rigopoulou et al.\,(2006). In Fig. \ref{fig:ms}, we present the position of the two sources in the SFR-$M_{\ast}$ plane along with the stellar mass and redshift 
dependent MS description of Schreiber et al.\,(2015) at $z=3.0$. Despite their high \lir\ both sources appear to have specific star formation rates consistent with that of Main sequence galaxies at their corresponding redshift and are placed at the high-mass end of the relation, where there is recent evidence for a bending of the MS (e.g. Schreiber et al.\,2015., Lee et al.\,2015). 

Finally, using the extinction values from the best fit BC03 models (E(B$-$V) $\approx$ 0.27 for both sources) we derive UV dust-corrected SFRs ($SFR_{\rm UVcor}$) estimates,  by converting the extinction-corrected UV-luminosity $L_{1500}$ to SFR (Daddi et al.\,2007, Magdis et al.\,2010c). For D49, we find $SFR_{\rm UVcor} = 325$\,\msol\,yr$^{-1}$ and for M28, $SFR_{\rm UVcor} = 230$\,\msol\,yr$^{-1}$. While these UV-based SFR estimates are $\sim$$1.5-2$ times lower compared to the IR-based SFR, this discrepancy is considerably lower than what is found for local ULIRGs and high$-z$ starbursts ($SFR_{\rm IR}/SFR_{\rm UV cor} >>$ 3), and closer to that found for normal galaxies (e.g., Elbaz et al. 2007; Magdis et al. 2010c; Rigopoulou et al. 2010).
 
\section{Discussion}
\subsection{Evolution of the mean radiation field $\langle U \rangle$}

The dust mass weighted luminosity of the galaxies (\md/\lir), (proportional to the mean radiation field, \um\ $\propto$ \md/\lir)  and its evolution with redshift, has been shown to  trace the evolution of the shape of the far-IR SED, of the dust temperature of the ISM and more recently of the gas phase metallicity in MS galaxies along cosmic time. In particular, this quantity yields  the emitted energy per unit dust mass, and for the case of a MBB with  an effective dust emissivity index of $\beta$, it is related to \td\ through the following formula: 

\begin{equation}
\begin{centering}
{\rm \langle U \rangle} \propto  {\rm \frac{L_{IR} }{ M_{\rm dust}} } \propto T_{d}^{4+\beta}
\end{centering}
\end{equation}

\noindent Using stacked ensembles of MS galaxies, Magdis et al.\,(2012b) and Bethermin et al.\,(2015), showed that for galaxies within the MS \um\  increases with redshift  as $(1+z)^{\zeta}$ (with $\zeta$ ranging between 1.2 and 1.8 among several studies) providing evidence of  MS galaxies becoming warmer as we move back in time. A similar evolution of the \td\ of MS galaxies with redshift has also been reported by Magnelli et al.\,(2014).

Furthermore, the evolution of \um\ with redshift can be used to test various scenarios of the evolution of the gas-phase metallicity of the galaxies. In particular, it is still debated if the fundamental metallicity relation (FMR, Mannucci et al.\,2010) that relates the gas phase metallicity to the \ms\ and the SFR of a star forming galaxy, holds beyond $z = 2$, or whether it breaks down, with galaxies at higher$-z$ having lower metallicities than those inferred by the FMR prescription. The latter is supported by various recent studies (e.g., Troncoso et al.\,2014, Amorin et al.\,2014, Steidel et al.\,2014) including  Onodera et.\,(2016), who showed that the metallicity of $z \sim  3.0$ star forming galaxies is offset by $\sim 0.3$\,dex from the locally defined FMR relation. Their  results are consistent with the redshift evolution of the mass-metalicity relation prescribed  by Lilly et al.\,(2013).

In Magdis et al.\,(2012b) we showed that since \um\ $\propto$ \lir/\md, \md\ $\propto$ \mgas\ $\times$ Z(\ms,SFR) (e.g., Leroy et al.\,2011) and \lir\ $\propto$ SFR $\propto$ $M^{\rm \gamma}_{\rm gas}$ (Kennicutt relation) then:

\begin{equation}
\begin{centering}
{\rm \langle U \rangle} \propto  {\rm \frac{L_{IR}}{M_{gas} \times Z}} = {\rm \frac{SFR^{1-\gamma}}{Z(M_{\ast},SFR)}}  
\end{centering}
\end{equation}  

Assuming $\gamma=0.83$, as reported by Sargent et al.\,(2014), Bethermin et al.\,(2015), calculated the expected evolution of \um, for the following two evolutionary scenarios for the gas phase metallicity: 1) a redshift invariant FMR (universal FMR) and 2) an FMR relation with a correction of 0.30$\times (1.7-z)$\.dex (broken FMR), to account for the deviation between recent metallicity measurements of high$-z$ galaxies and those prescribed by a universal FMR (see Tan et al.\,2014). At fixed stellar mass, the first case predicts a smooth increase of  \um\ with redshift, depicting the decrease of Z as we move to higher$-z$, due to the increase of SFR with redshift along the MS. On the other hand, a broken FMR predicts a steep increase of \um\ beyond $z>1.7$. Comparing the two evolutionary tracks to the average \um\ values of MS galaxies at different redshift bins as derived from stacking, they found that the observed evolution of \um, is consistent with the case of a broken FMR. 

\begin{figure}
\centering
\includegraphics[scale=0.42]{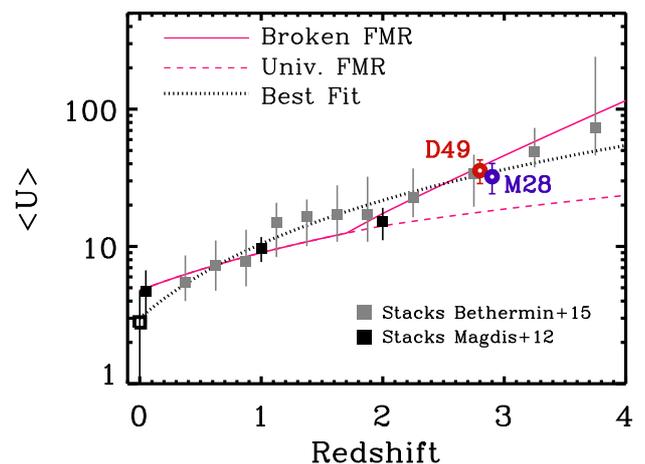}
\caption{The evolution of the mean radiation field, or equally, of the dust mass weighted luminosity, with redshift of main sequence galaxies. The red and blue circles correspond to the $\langle$U$\rangle$ values of D49 and M28 as derived from the DL07 models. The grey and black squares correspond to the average \um\ values of MS galaxies as derived from the stacking analysis of Bethermin et al.\,(2015) and Magdis et al.\,(2012b) respectively. The open black square corresponds to the mean value of the local Herschel Reference Sample by  Ciesla et al.\,(2013). The solid and dashed pink lines represent the evolutionary trends expected for a broken and universal FMR, respectively, while the dotted line depicts the best fit to the data.}
\label{fig:u} %
\end{figure}

Here we can further inform and test the observed trends emerging from stacking results with direct measurements of \lir\ and \md\ of individual MS galaxies at $z \sim 3$. In Figure \ref{fig:u}, we present the evolution of \um\ with redshift as derived from stacking of MS galaxies at 
various redshift from Magdis et al.\,(2012b) and Bethermin et al.\,(2015), along with the  direct measurements for our two galaxies. With \um\ measurements of 35.7 $\pm$ 7.0 for D49 and 32.2 $\pm$ 8.0 for M28, the individual detections appear to be in excellent agreement with the average  \um\ at 
the corresponding redshift presented in Bethermin et al.\,(2015). Also, our measurements appear consistent with a broken FMR, lying $\sim2-3\sigma$ above the prediction of a universal FMR. Brought together our analysis, confirms the increase of \um\ of MS galaxies at $z>2$, and supports 
the scenario where the metallicity of $z>2$ galaxies is better described by a broken FMR.

\subsection{Molecular gas mass estimates}
In recent years we have witnessed great progress in the determination of the gas mass reservoirs of galaxies across cosmic time. Such 
measurements are based  either 1) on the more traditional conversion of the integrated CO line luminosity \lco\ to \mol\ using the $
\alpha_{CO}$ conversion factor, or 2) on the more recently introduced metallicity-dependent dust-to-gas mass ratio technique ($\delta_{\rm 
GD}$), that exploits  the consistent/accurate derivation of \md\ allowed by detailed sampling of the far-IR SED by \h\ and other (sub)mm 
facilities or 3) on the  single band measurement of the dust emission flux on the Rayleigh-Jeans side of the SED (e.g. Scoville et al.\,2014, Groves et al.\,2015, Schinnerer et al.\,2016).

With our data we are now able to compare measurements of \mol\ inferred using all three independent methods for the first time for $z = 3$ MS galaxies. 

\subsubsection{$CO$ based $M_{\rm H_{2}}$}
The \mol\ of the galaxies is related to the \cooz\ line luminosity (\lco) through the equation:   
\begin{equation}
\begin{centering}
M_{\rm H_{\rm 2}}\,[M_{\odot}] =  \alpha_{\rm CO} \times L^{\prime}_{\rm CO},
\end{centering}
\end{equation}
\noindent where $\alpha_{\rm CO}$ is the CO to \mol\ conversion factor.   

Since our PdBI  observation of D49 traces the \cott\ rather than the fundamental \cooz\ emission line  we need to adopt an excitation correction ($r_{\rm 31}$) to convert the observed $L_{\rm CO[3-2]}$ to $L^{\prime}_{\rm CO}$. While, Carilli \& Walter\,(2013) find that $r_{\rm 31}$  can vary from 0.27 (Milky Way) to 0.97 (quasars), studies of typical star 
forming galaxies and starburst galaxies find a fairly consistent average value of  $\langle\,r_{\rm 31}\,\rangle \sim 0.42-0.65$ 
(Dannerbauer et al.\,2009, Tacconi et al.\,2010, Carilli \& Walter\,2013; Greve et al. 2014; Genzel et al.\,2015, Daddi et al.\,2015, Sharon et 
al.\,2016). Here, we adopt the median value of $\langle\,r_{\rm 31}\,\rangle \sim 0.5$ reported in various studies, along with an uncertainty 
of 0.15, to account for the observed variations among various samples in the literature, and derive $L^{\prime}_{\rm CO} = (8.67 \pm 2.6) 
\times 10^{10}$\,\lsol.

In addition to the adopted $r_{\rm 31}$, a further assumption needs to be made regarding the choice of the value of $\alpha_{\rm CO}$, that is known to vary as a function of metallicity and possibly depends on the star formation mode of the galaxy. Since we lack direct measurements of the gas-phase metallicity for D49, we choose to adopt a solar metallicity that is very close to the average value between the cases of a broken and universal FMR discussed above. Subsequently, we use various \aco\ $- Z$ relations reported in the literature (Leroy et al.\,2011, Magdis et al.\,2012b, Genzel et al.\,2012) and derive an average \aco\ $= 3.5$, which is a typical value for MS galaxies at various redshifts.  Under these assumptions and combining all uncertainties in quadrature we derive $log($\mol\,[\msol]$)$ $= 11.48 \pm 0.23$.
\begin{figure*}
\centering
\includegraphics[scale=0.35]{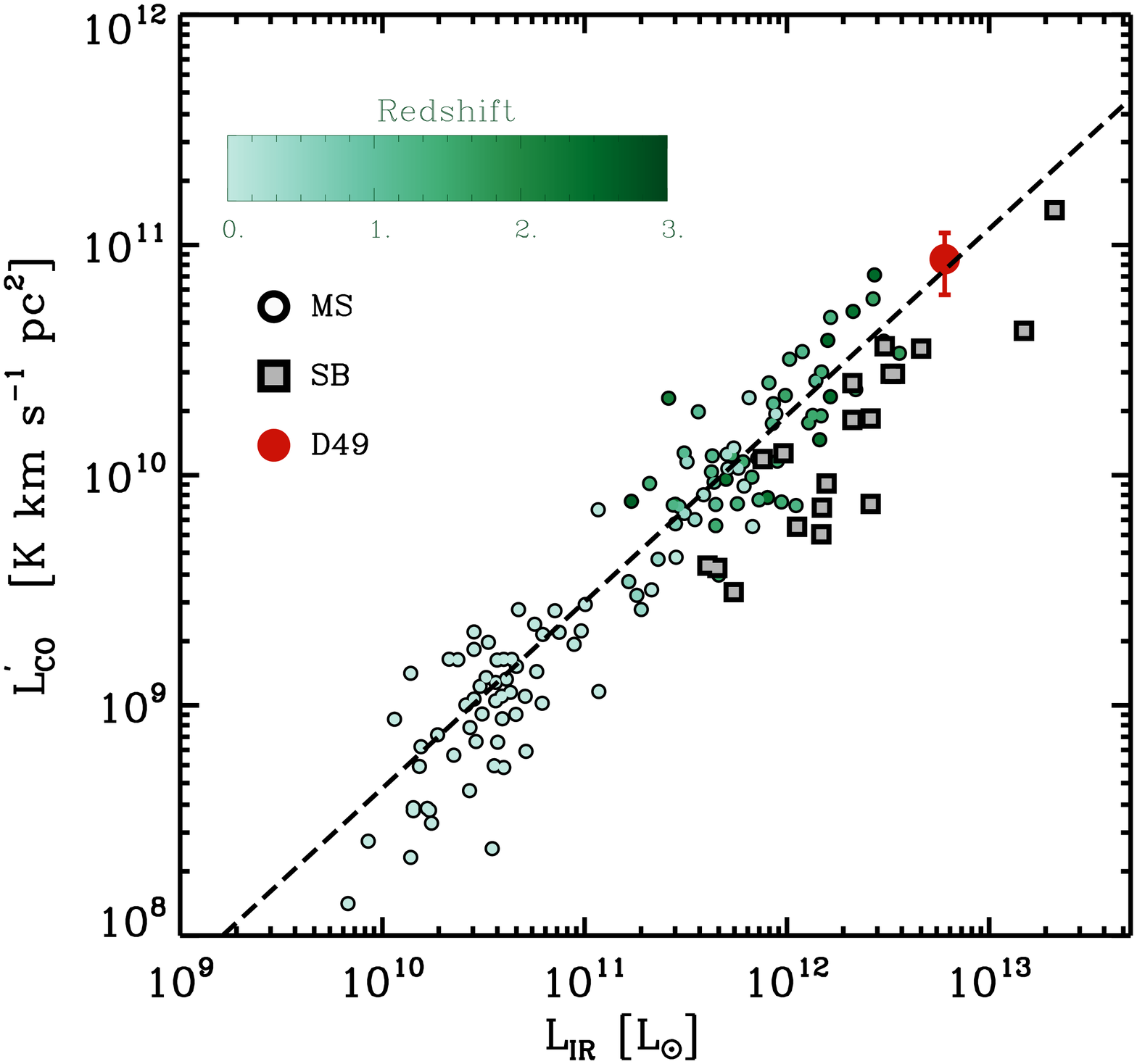}
\includegraphics[scale=0.35]{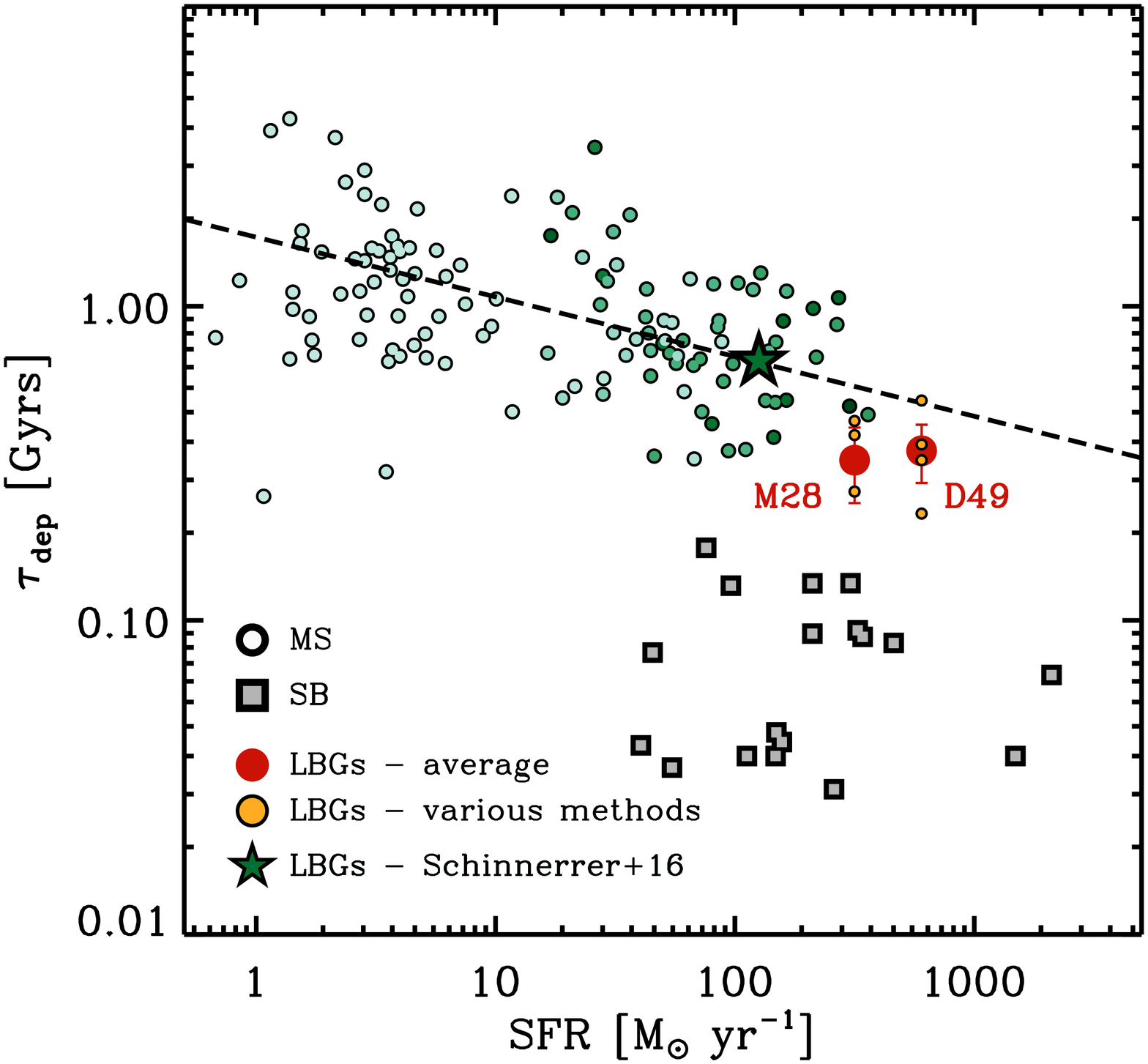}
\caption{\textbf{Left:} Integrated Schmidt-Kennicutt relation using direct observables. Green circles correspond to MS galaxies, colour coded according to their redshift. Grey squares correspond to local and high$-z$ star-bursts galaxies ($\times4$ above the MS). The compilation of MS and SBs galaxies is presented in Sargent et al.\,(2014) and Silverman et al.\,(2016).The red dot depicts the position of D49, after converting the measured \cott\ luminosity to  \cooz\, assuming a line ratio of $r_{\rm 3,1} = 0.5$. The dashed line corresponds to the integrated Schmidt-Kennicutt relation of MS galaxies as derived by Sargent et al.\,(2014). \textbf{Right:} Gas depletion time scale ($\tau_{\rm dep}$ = \mol/SFR or 1/SFE), based on the average \mol\ estimates from the various methods described in the text (filled red circles). The individual measurements based on various techniques and metallicity assumptions for each galaxy are depicted with orange circles. As in the left panel, literature MS galaxies are colour coded based on their redshift (see Sargent et al.\,2014 for more details about the sample compilation). The dark green star corresponds to the average measurement reported by Schinnerer et al.\,(2016) for a sample of typical $z\sim3.2$ LBGs, derived based on single mm band observations. The dashed line corresponds to the sequence of MS galaxies as derived by Sargent et al.\,(2014).}
\label{fig:gas} %
\end{figure*}

\subsubsection{$\delta_{\rm GD}$ based $M_{\rm H_{2}}$}
This method and its associated uncertainties have been presented in detail in various studies (e.g., Magdis et al.\,2011,2012b, Berta et al.\,2016). In brief, if the metallicity and the \md\ of a galaxy are known, then one can use the well calibrated local \mgas/\md\ vs Z relation to derive \mgas, and under the assumption that the former does not evolve considerably with look back time. As pointed out in Magdis et.\,(2012b) and in more detail by Berta et al.\,(2015), a critical component of this method relies on the existence of photometric data both at the peak of the far-IR SED but also in the  R$-$J tail ($\lambda_{\rm rest} > 250\,\mu$m). We note that this technique traces the total amount of neutral hydrogen in the ISM, thus \mgas\ $= M_{\rm H_{\rm 2}} +M_{\rm HI}$. However,  based on observational evidence (e.g., Daddi et al. 2010a; Tacconi et al.\,2010; Geach et al.\,2011) as well as on theoretical arguments (Blitz \& Rosolowsky\,2006, Bigiel et al.\,2008, Obreschkow et al.\,2009) we can assume that for our $z\sim3$ galaxies, $M_{\rm H_{\rm 2}} >>M_{\rm HI}$  or equivalently that \mgas\ $\approx M_{\rm H_{\rm 2}}$.

Using the \md\ estimates from the DL07 models and assuming a solar metallicity for both sources we derive a gas mass of  $log($\mol\,[\msol]$)$ $= 11.12 \pm 0.25$ for D49 and $log($\mol\,[\msol]$)$ $= 10.91 \pm 0.31$ for M28. Repeating the analysis based on metallicities derived from a broken FMR, would yield higher gas masses by a factor of $\sim$1.5 for both sources, i.e. $11.34 \pm 0.25$ for D49 and  $11.12 \pm 0.31$ for M28. 

\subsubsection{$R-J$ based $M_{\rm H_{2}}$}  
 The available 1.2\,mm data, that for the redshift of our sources correspond to $\lambda_{\rm rest} \approx 300\,\mu$m, allows for a third estimate of \mgas\ that relies on single band measurement of the dust emission flux on the Rayleigh-Jeans side of the SED (e.g., Scoville et al.\,2014,2016). This technique converts the observed mm flux density into a cold gas mass, by making use of the observed relation between cold dust luminosity and gas mass in nearby galaxies.   
 
 Here we consider the calibration presented in Schinnerer et al.\,(2016), that was applied to a sample of $z\sim3$ LBGs with available ALMA data in the R-J tail of the SED.   Using the relations between monochromatic IR luminosities at 250, 350 and 500\,$\mu$m and neutral gas mass from Groves et al.\,(2015) that are calibrated on observations for 36 local galaxies from the KINGFISH survey (Kennicutt et al.\,2011),  and adding a linear correction to the coefficients to account for the difference between observed and rest-frame wavelengths Schinnerer et al.\,(2016)  provide the following formula: 
 
 \begin{equation}
\begin{centering}
\begin{split}
log(M_{\rm gas} [M_{\odot}]) =  (1.57 - 8 \times 10^{-4} \Delta\lambda) \\
+(0.86+6 \times 10^{-4} \Delta\lambda)\\
\times\, log(\nu L_{\nu} [L_{\odot}])
\end{split}
\end{centering}
\end{equation}
 
\noindent where 
  \begin{equation}
\begin{centering}
\begin{split}
\Delta\lambda = \lambda_{\rm rest} - 250\,\mu m\,\, \rm {and}\,\, \nu L_{\nu} = \nu_{\rm obs} \times S_{\rm \nu,obs} \times 4\pi \times D^{2}_{\rm L}
\end{split}
\end{centering}
\end{equation}
\noindent Using the observed flux densities ($S_{\rm \nu,obs}$) at 1.2\,mm we infer (assuming again that \mgas\ $\approx M_{\rm H_{\rm 2}}$) $log(M_{\rm H_{\rm 2}}$\,[\msol]$)$ $= 11.29 \pm 0.31$ for D49 and  $log($\mol\,[\msol]$)$ $= 11.10 \pm 0.31$ for M28, in excellent agreement with the values derived based on the $\delta_{\rm GD}$ approach. The limitations of this technique, i.e. not taking into account the evolution of the gas-phase metallicity and of the dust temperature of the galaxies (at fixed stellar mass) with redshift, are discussed in detail in Genzel et al.\,(2015), Berta et al.\,(2016) and Schinnerer et al.\,(2016). These limitations are more severe for the case of high$-z$ star-bursting systems whose gas-phase metallicity could deviate much more than that of MS galaxies with respect to  calibrations in the local universe. Thus, extra caution should be exercised when this technique is applied in high$-z$ SB systems.

Comparing the \mol\ estimates from the various techniques we find an overall agreement within a factor of $\sim$ 2 or less, and \mol\ values that are consistent within the uncertainties linked wth each approach (Table 3).

\subsection{Star formation law and SFE of $z=3$ MS galaxies}
  
Several studies have provided evidence that main sequence star-forming galaxies at all redshifts appear to follow a tight relation between their gas mass reservoir and their star formation rate, commonly referred to as the integrated Schmidt-Kennicutt relation. However, the sensitivity limits of \h\ and the time demanding nature of CO observations for MS galaxies, has largely restricted the measurement of the gas content of  $z \sim 3$ MS galaxies to lensed objects or stacked ensembles (e.g., Saintonge et al.\,2013), or more recently to  \mol\ measurements based on a single band measurement of the dust emission flux on the Rayleigh-Jeans side of the SED (e.g. Scoville et al. 2014, Schinnerer et al. 2016). Indeed, direct CO detections of unlensed LBGs at $z>3$ are restricted to two sources (e.g. Magdis et al.\,2012a, Tan et al.\,2014). Here, with three independent \mol\ measurements we are in position to investigate the star formation efficiency, defined by the ratio of the star formation rate over the molecular gas mass (SFE=SFR/\mol) of MS galaxies at $z \approx 3$. 

Since \lir\ $\propto$ SFR and \lco\ $\propto$ \mol, this  implies that SFE $\propto$ \lir/\lco. Therefore, it is useful to first consider the direct observables, \lir\ and \lco, 
that are subject to fewer assumptions and uncertainties, before using the inferred SFRs and gas masses to investigate the SFE at $ z\sim3$. In Fig. 
\ref{fig:gas}(left), we compare the CO and IR luminosity of D49, with other cosmologically relevant populations of MS and SB galaxies, drawn from the compilation 
presented in Sargent et al.\,(2014) along with data from Magdis et al.\,(2014) and Silverman et al.\,(2015). When required, higher transition CO luminosities (\coto\ and \cott), are converted to \lco, assuming  typical excitation corrections ($r_{\rm 21}=0.85$  and $r_{\rm 31}=0.5$).
It is clear that D49, follows the trend of local and high$-z$ MS galaxies, as defined by Sargent et al.\,(2014) with a \lir/\lco\ ratio of (71 $\pm$ 17)\,\lsol/[K\,km\,s$^{-1}$\,pc$^{2}$]. While whether (a) SBs deviate from the \lir$-$\lco\ relation of MS galaxies as a power-law to their offset from the MS (e.g. Tacconi et al.\,2017, Scoville et al.\,2017), or (b) there is a more abrupt transition between the MS and SB regime (2-SFM paradigm, Sargent et al.\,2014), is still an open question, it is clear that SB galaxies have systematically larger \lir/\lco ratios than MS galaxies. The average  $\langle$\lir/\lco $\rangle$  for SBs is 160\,\lsol/[K\,km\,s$^{-1}$\,pc$^{2}$], indicating a higher star formation efficiency with respect to D49 and other MS systems at various redshift by a factor of $\sim$2 or more, just by considering direct observables. 

Using the derived SFR and \mol\ estimates of the two galaxies studied here, we then infer their gas depletion time scales defined as $\tau_{\rm dep} =$ 1/SFE. For each of the two galaxies, instead of choosing a single \mol\ measurement from one of the methods described above and summarised in Table 3, we adopt  the average \mol\ as derived from all  three  (two) independent methods for D49 (M28). By combining the SFRs and the \mol\ of our galaxies, we find that both sources are characterised by very similar gas depletion time scales with $\tau_{\rm dep} = (0.35 \pm 0.13)\,Gyr$ and $\tau_{\rm dep} = (0.32 \pm 0.12)\,Gyr$ for D49 and M28 respectively. This is a factor of $\sim2$ lower compared to the average $\tau_{\rm dep}$ found by Schinnerer et al.\,(2016) for a sample of less massive $z\sim3$ LBGs. However, the $\tau_{\rm dep}$ of our sources  lie within the scatter of the $\tau_{\rm dep} -$ SFR relation of Sargent et al.\,(2014) defined by MS galaxies across various cosmic epochs (Fig.\ref{fig:gas},right). 

Moving to gas-to-stellar mass ratios ($f_{\rm gas} =$ \mol/$M_{\ast}$), we find $f_{\rm gas} =1.1\pm0.2$ for D49 and $0.4\pm0.1$ for M28. Both values are consistent with previous studies that provide evidence for an increasing \fgas\ in MS galaxies with look-back time, that at $z>2$ reaches a factor of $\sim5-10$ with respect to local spirals  (e.g. Magdis et al.\,2012a,b, Geach et al.\,2011, Tan et al.\,2014). The factor of $\sim$2.5 difference in \fgas\ between the two galaxies in our sample should not come as surprise; apart from the expected intrinsic scatter, several studies have shown that  \fgas\ among MS galaxies at a given redshift, increases as a power of the distance to the MS, $(sSFR/sSFR_{\rm MS})^{\gamma}$ (e.g. Magdis et al.\,2012b, Sargent et al.\,2014, Genzel et al.\,2015, Scoville et al.\,2017). Adopting $\gamma=0.8$ from Magdis et al.\,(2012b) and taking into account that D49 and M28 lie $\times$1.5  and $\times$1.4 above and below the MS respectively, suggests $\sim\times$2 higher \fgas\ in D49 compared to M28. Thus, the observed difference in \fgas\  between the two  galaxies is fully consistent with the notion of varying \fgas\ among MS galaxies at a given redshift. We note that we reach a similar conclusion  when we consider a weaker dependance of \fgas\ on  $sSFR/sSFR_{\rm MS}$ ($\gamma \approx 0.3-0.6$) reported by other studies (e.g. Genzel et al.\,2015, Scoville et al.\,2017, Tacconi et al.\,2017).

Put together, our analysis suggests that individual $z\sim3$  MS galaxies appear to follow the SFR\ $-$ \mol\ and \lir\ $-$ \lco\ relations that are established at lower redshifts, indicating that overall MS framework holds at least up to $z\sim3$.

\section{Conclusions}
Combining CO and dust measurements,  we perform a detailed analysis of the dust and gas properties of two $z\sim3$ MS galaxies. These are among a handfull of individually detected, unlensed, ``normal'' star forming galaxies at $z>2.5$ in the literature, with both CO detections and fully characterised  IR SEDs out to the  mm bands.  Our findings can be summarised as follows:

\begin{itemize}
\item We confirm the increase of the mean radiation field \um\ as a function of redshift, at least up to $z=3$. This increase of \um\ can be interpreted as a signature of an evolution toward lower gas-phase metallicities in $z > 2$ galaxies that is increased (``broken'' FMR scenario in Sect. 4.1) with respect to expectations based on the universal FMR.

\item For high$-z$ MS galaxies, the various techniques to derive \mol\  provide consistent  measurements, within a factor of $\sim$2 (or less).  We note however, that this might not be the case for SBs, as  their metallicity remains largely unconstrained.

\item Our galaxies appear to follow the \lir\ $-$ \lco, SFR$-$\mol\ relation established by ``normal" galaxies at lower redshifts, extending the apparent uniformity of the star forming galaxies out to $z=3$. 

\end{itemize}

\textit{Acknowledgments.} This work is based on observations carried out under project number S14CN002, with the IRAM PdBI Interferometer. IRAM is supported by INSU/CNRS (France), MPG (Germany) and IGN (Spain). We are grateful to Prof. Alice Shapley for providing the UV spectrum of D49.
GEM acknowledges support from the Carlsberg Foundation, the ERC Consolidator Grant funding scheme (project ConTExt, grant number No. 648179), and a research grant (13160) from Villum Fonden. DR acknowledges support from ST/K00106X/1 and ST/N000919/1. CF acknowledges funding from
the European Union's Horizon 2020 research and innovation programme under the Marie Sklodowska-Curie Grant
agreement No 664931. MTS acknowledges support from a Royal Society Leverhulme Trust Senior Research Fellowship (LT150041). H.D. acknowledges financial support from the Spanish Ministry of Economy and Competitiveness (MINECO) under the 2014 Ramon y Cajal program MINECO RYC-2014-15686.

\bibliographystyle{aa} % style aa.bst
\bibliography{ms_v3}

\onecolumn
\begin{deluxetable}{lccccccccc}

\tabletypesize{\normalsize}
\tablewidth{0pc}
\tablecaption{Mid to Far-IR and mm Photometry}
\tablehead{

\colhead{Name} &
\colhead{$z_{\rm spec}$} &
\colhead{$S_{\rm 24}$} &
\colhead{$S_{\rm 100}$} &
\colhead{$S_{\rm 160}$} &
\colhead{$S_{\rm 250}$} &
\colhead{$S_{\rm 350}$} &
\colhead{$S_{\rm 500}$} &
\colhead{$S_{\rm 1.2}$}\\

\colhead{} &
\colhead{} &
\colhead{[mJy]} &
\colhead{[mJy]}&
\colhead{[mJy]}&
\colhead{[mJy]}&
\colhead{[mJy]} &
\colhead{[mJy]}&
\colhead{[mJy]}&} 
\startdata
 D49  &   2.846  &    0.09$\pm$0.03&  4.27$\pm$1.22 & 12.21$\pm$3.56&15.08$\pm$1.80&18.19$\pm$1.84&11.83$\pm$2.37&1.79$\pm$0.37 \\
 M28  &   2.908  &    0.18$\pm$0.02&  2.05$\pm$1.27 & 4.33$\pm$2.8&10.12$\pm$2.21&13.75$\pm$1.87&5.32$\pm$2.19&1.10$\pm$0.36 \\
 \enddata
\end{deluxetable}

\begin{deluxetable}{lccccccccc}

\tabletypesize{\normalsize}
\tablewidth{0pc}
\tablecaption{Physical Properties}
\tablehead{

\colhead{Name} &
\colhead{log$L_{\rm IR}$} &
\colhead{$I_{CO[3-2]}$}&
\colhead{$L^{\prime}_{CO}$ $^{a}$}&
\colhead{log$M_{\rm dust}$ $^{b}$} &
\colhead{log$M_{\rm H_{2}}$ $^{c}$} &
\colhead{log$M_{\ast}$} &
\colhead{$\langle U \rangle$} &
\colhead{$T_{\rm d}$ $^{d}$}& \\

\colhead{} &
\colhead{[$L_{\odot}$]} &
\colhead{[Jy\,km\,sec$^{-1}$]} &
\colhead{[K\,km\,sec$^{-1}$\, pc$^{2}$]}&
\colhead{[$M_{\odot}$]}&
\colhead{[$M_{\odot}$]}&
\colhead{[$M_{\odot}$]}&
\colhead{-}&
\colhead{[K]}} 
\startdata
D49    &  12.78$\pm$0.03& 1.10$\pm$0.18&10.93$\pm$0.11&9.12$\pm$0.12&$11.32\pm$0.15&11.28$\pm$0.12&$35.7\pm$7.0&40.6$\pm$2.0\\
M28    &  12.51$\pm$0.04& - & - &8.90$\pm$0.15&$11.01\pm$0.12&11.38$\pm$0.10&$32.2\pm$8.0&41.8$\pm$3.0\\

 \enddata
 
 Notes: \\
a: CO[1-0] luminosity assuming an excitation correction of $r_{\rm 31} = 0.5$. The quoted uncertainty includes  the observed uncertainty in the flux measurement and the uncertainty in $r_{\rm 31}$, in quadrature. \\
b: Derived based on DL07 models.\\
c: Average value, between CO, $\delta_{\rm GD}$ and R$-$J approach.\\
d: Based on a MBB model with fixed $\beta = 1.8$ 
 \end{deluxetable}

\begin{deluxetable}{lcccc}
\tabletypesize{\normalsize}
\tablewidth{0pc}
\tablecaption{Molecular gas mass  estimates}
\tablehead{

\colhead{Name} &
\colhead{CO} &
\colhead{$\delta_{GD}$ $Z_{\odot}$}&
\colhead{$\delta_{GD}$ ``\it{broken}'' FMR}&
\colhead{R$-$J} \\
\colhead{} &
\colhead{log$(M_{\rm H_{2}}/M_{\odot})$}&
\colhead{log$(M_{\rm H_{2}}/M_{\odot})$}&
\colhead{log$(M_{\rm H_{2}}/M_{\odot})$}&
\colhead{log$(M_{\rm H_{2}}/M_{\odot})$}} 

\startdata

D49    &  11.48$\pm$0.23 & 11.12$\pm$0.25 &11.34$\pm$0.25 &11.29$\pm$0.31\\
M28    &  - & 10.91$\pm$0.31                        & 11.12$\pm$0.31                        &11.10$\pm$0.34\\

\enddata  
\end{deluxetable}

{}

\end{document}